\documentclass[12pt,preprint]{aastex}

\shorttitle{Heating by Cosmic-Ray Streaming}
\shortauthors{Fujita \& Ohira}

\begin{document}

\title{Stable Heating of Cluster Cooling Flows by Cosmic-Ray Streaming}

\author{Yutaka Fujita}
\affil{Department of Earth and Space Science, Graduate School of
Science, Osaka University, 1-1 Machikaneyama-cho, Toyonaka, Osaka
560-0043, Japan}
\email{fujita@vega.ess.sci.osaka-u.ac.jp}

\and

\author{Yutaka Ohira}
\affil{Theory Centre, Institute of Particle and Nuclear 
Studies, KEK, 1-1 Oho, Tsukuba 305-0801, Japan}

\begin{abstract}
 We study heating of cool cores in galaxy clusters by cosmic-ray (CR)
 streaming using numerical simulations. In this model, CRs are injected
 by the central active galactic nucleus (AGN) and move outward with
 Alfv\'en waves. The waves are excited by the streaming itself and
 become non-linear. If magnetic fields are large enough, CRs can prevail
 in and heat the entire core because of a large Alfv\'en velocity. We
 find that the CR streaming can stably heat both high and low
 temperature clusters for a long time without the assistance of thermal
 conduction, and it can prevent the development of massive cooling
 flows. If there is even minor contribution of thermal conduction, the
 heating can be more stabilized. We discuss the reason of the stability
 and indicate that the CR pressure is insensitive to the change of
 intracluster medium (ICM), and that the density dependence of the
 heating term is similar to that of the radiative cooling.
\end{abstract}


\section{Introduction}

Clusters of galaxies are filled with hot X-ray gas (intracluster medium;
ICM). Although the cooling time of the ICM is larger than the age of the
Universe for the most part of a cluster, the core is the exception
\citep{sar86}. Since the cooling time at the core ($\sim 10^8$~yr) is
much smaller than the age of the cluster ($\gtrsim 10^9$~yr), a
substantial gas inflow, which was called a ``cooling flow'', was
expected to develop (\citealt{fab94}, and references therein). However,
X-ray spectra taken with {\it ASCA} and {\it XMM-Newton} did not detect
line emission from intermediate or low temperature gas
\citep[e.g.][]{ike97,mak01,pet01,tam01,kaa01,mat02}. This means that the
the cooling rate is much smaller than that previously assumed. {\it
Chandra} observations are also consistent with these results
\citep*[e.g.][]{mcn00,joh02,ett02,bla03}. These observations show that
some heat sources balance with radiative cooling and prevent the
development of cooling flows.

Many mechanisms for heat sources of the ICM have been proposed. Thermal
conduction from the hot outer layers of clusters is one popular idea
\citep{tak79,tak81,tuc83}. This model may work for clusters with a
middle or high temperature if the conductivity is $\sim 30$\% of the
Spitzer one \citep{zak03}. However, the conductivity requires
fine-tuning \citep{bre88,bri03,sok03,guo08}. Moreover, since the Spitzer
conductivity significantly decreases for low temperature clusters,
thermal conduction may not be able to transfer enough energy. The active
galactic nuclei (AGNs) at the cluster centers are also considered as
heat sources \citep*[e.g.][]{chu01,qui01,bru02,bas03}. In fact, it has
been observed that the AGNs disturb the surrounding ICM
\citep*[e.g.][]{fab00,mcn00,bla01,mcn01,maz02,fuj02,joh02,kem02,tak03,fuj04}. However,
it is not clear how the energy generated by the AGN is transfered to the
ambient ICM. For example, sound waves observed around the AGNs in some
clusters were considered the significant channel for the AGN energy
input into the ICM \citep{fab03a,for05}. However, theoretical studies
have indicated that the dissipation of the sound waves or weak shocks is
too fast to heat an entire cool core and thus the waves cannot stably
heat the core (\citealp*{fuj05,mat06}; see also \citealp{fuj07b}).

Cosmic-rays (CRs) may be another channel for the AGN energy input into
the ICM \citep*[e.g.][]{tuc83,boh88,rep87,rep95,col04,pfr07,jub08}. In
particular, CR streaming has been investigated as a way through which
the CR energy is transfered to the ICM \citep*{boh88,loe91}. In this
mechanism, $PdV$ work done by the CRs on an Alfv\'en wave effectively
becomes an energy source of the ICM (see \S~\ref{sec:crst}). It is to be
noted that this mechanism may also be working in supernova remnants
\citep*[e.g.][]{voe84,ber99}.  For clusters of galaxies, the models of
\citet{boh88} and \citet{loe91} are time-independent. Thus, the models
cannot treat the ICM that is not in a steady state. Recently,
\citet{guo08} studied non-steady models. They combined CR streaming and
thermal conduction as heat sources, and showed that the ICM turns out to
be in a steady state, and that a cooling flow is suppressed. In their
models, CR streaming and thermal conduction seem to equally contribute
to the heating of the ICM (their Figure~6). Since they constructed
models for a middle-temperature cluster (A~2199), it is not certain
whether the models can be applied to low temperature clusters in which
thermal conduction is not effective \citep[see][]{bri03}. Moreover, the
conductivity of the ICM may be small even for hot clusters. The
existence of cold fronts in some clusters means that at least in some
regions in those clusters, the conductivity has to be much smaller than
the Spitzer value \citep{ett00}. Simultaneous heating by both the
central AGN and thermal conduction had been proposed by
\citet{rus02}. In their model, mechanical heating through bubble motion
was considered for the AGN heating. Since the advantage of CR streaming
over the mechanical heating was not clear in the models of
\citet{guo08}, it may be useful to study the case where a cool core is
heated only by CR streaming.

In this study, we revisit the heating of the ICM by CR streaming. We
consider non-steady ICM and investigate whether the CR streaming alone
can stably heat the ICM without the assistance of thermal conduction. We
emphasize that we do not intend to search models in which the heating is
completely balanced with radiative cooling. Since the age of clusters is
finite \citep[say $\sim 5$~Gyr; e.g.][]{kit96b}, it is sufficient to
find solutions that are stable for that time. We treat both high and low
temperature clusters. We follow the growth of Alfv\'en waves, which
\citet{guo08} did not consider. We simulate for various parameters in
order to find the nature of the CR heating, although our goal is not to
compare the results with specific clusters in detail, because
supplemental heating mechanisms other than CR streaming (e.g. shock
heating) are likely to be effective in real clusters. We consider
protons as CRs.

The paper is organized as follows. In \S~2, we explain our models for
CR streaming and galaxy clusters. In \S~3, we present the results for
various parameters, and show that the CR heating is fairly stable. In
\S~4, we discuss the reason of the stability. \S~5 is devoted to
conclusions.

\section{Models}

\subsection{Basic Equations}

For simplicity, we assume that the cluster is spherically symmetric.
The flow equations are
\begin{equation}
 \frac{\partial \rho}{\partial t} 
+ \frac{1}{r^2}\frac{\partial}{\partial r}(r^2\rho u) = 0\:,
\end{equation}
\begin{equation}
\frac{\partial (\rho u)}{\partial t} 
+ \frac{1}{r^2}\frac{\partial}{\partial r}(r^2\rho u^2)
= - \rho \frac{G M(r)}{r^2}-\frac{\partial}{\partial r}
(P_g + P_c + P_B)\:,
\end{equation}
\begin{eqnarray}
 \frac{\partial e_g}{\partial t}  
+ \frac{1}{r^2}\frac{\partial}{\partial r}(r^2 u e_g)
&=& -P_g \frac{1}{r^2}\frac{\partial}{\partial r}(r^2 u) 
+ \frac{1}{r^2}\frac{\partial}{\partial r}
\left[r^2\kappa(T)\frac{\partial T}{\partial r}\right]\nonumber\\
& &- n_e^2\Lambda(T) + H_{\rm st} + H_{\rm coll}\:,
\label{eq:eg}
\end{eqnarray}
\begin{equation}
\label{eq:ec}
 \frac{\partial e_c}{\partial t}  
+ \frac{1}{r^2}\frac{\partial}{\partial r}(r^2 \tilde{u} e_c)
= -P_c \frac{1}{r^2}\frac{\partial}{\partial r}(r^2 \tilde{u}) 
+ \frac{1}{r^2}\frac{\partial}{\partial r}
\left[r^2 D(\rho)\frac{\partial e_c}{\partial r}\right] 
- \Gamma_{\rm loss}
+ \dot{S}_c \:,
\end{equation}
where $\rho$ is the gas density, $u$ is the gas velocity, $P_g$ is the
gas pressure, $P_c$ is the CR pressure, $P_B$ is the magnetic pressure,
$G$ is the gravitational constant, $M(r)$ is the gravitational mass
within the radius $r$, $\kappa(T)=f_c\kappa_0 T^{5/2}$ is the
coefficient for thermal conduction and $T$ is the temperature, $n_e$ is
the electron density, $\Lambda$ is the cooling function, $H_{\rm st}$ is
the heating by CR streaming, $H_{\rm coll}$ is the heating by Coulomb
and hadronic collisions, $\tilde{u}$ is the CR transport velocity,
$D(\rho)$ is the diffusion coefficient for CRs averaged over the CR
spectrum, $\Gamma_{\rm loss}$ is the energy loss by Coulomb and hadronic
collisions, and $\dot{S}_c$ is the source term of CRs. Energy densities
of the gas and the CRs are respectively defined as
$e_g=P_g/(\gamma_g-1)$ and $e_c=P_c/(\gamma_c-1)$, where $\gamma_g=5/3$
and $\gamma_c=4/3$.

For thermal conductivity, $\kappa_0=5\times 10^{-7}$ in cgs units
corresponds to the classical Spitzer value, and $f_c$ is the ratio to
that. The cooling function is based on the detailed calculations by
\citet{sut93},
\begin{equation}
\label{eq:cool}
 n_e^2 \Lambda = [C_1 (k_B T)^\alpha + C_2 (k_B T)^\beta + C_3]n_i n_e\:,
\end{equation}
where $n_i$ is the ion number density and the units for $k_B T$ are
keV. For an average metallicity $Z=0.3\: Z_\sun$ the constants in
equation (\ref{eq:cool}) are $\alpha=-1.7$, $\beta=0.5$, $C_1=8.6\times
10^{-3}$, $C_2=5.8\times 10^{-2}$, and $C_3=6.4\times 10^{-2}$, and we
can approximate $n_i n_e=0.704(\rho/m_H)^2$. The units of $\Lambda$ are
$10^{-22}\:\rm ergs\: cm^3$ \citep{rus02}. For Coulomb and hadronic
collisions, we use $H_{\rm coll}=\eta_c n_e e_c$ and $\Gamma_{\rm
loss}=\zeta_c n_e e_c$, where $\eta_c=2.63\times 10^{-16}\rm\: cm^3\:
s^{-1}$ and $\zeta_c=7.51\times 10^{-16}\rm\: cm^3\: s^{-1}$
\citep{guo08}. Since their contributions are minor, the details of the
collisions do not affect the results.

\subsection{CR Streaming and Heating}
\label{sec:crst}

Alfv\'en waves, which scatter CRs, are driven by CR streaming because
CRs can give their momentum to the waves through resonance
\citep[streaming instability;][]{ski75,bel78}. Since the CRs as a whole
move with Alfv\'en waves, the CR transport velocity in
equation~(\ref{eq:ec}) is given by $\tilde{u}=u+v_A$, where
$v_A=B/\sqrt{4\pi\rho}$ is the Alfv\'en velocity for a magnetic field
$B$ \citep*[e.g.][]{kan06,guo08,cap09b}. The wave energy $U_A=\delta
B^2/(8\pi)$, where $\delta B$ is the magnetic field fluctuation, is
amplified by the $PdV$ work done by the CRs on Alfv\'en waves:
\begin{equation}
\label{eq:UA}
 \frac{\partial U_A}{\partial t} 
= v_A\left|\frac{\partial P_c}{\partial r}\right|
\end{equation}
\citep{luc00}. We solve this equation by setting $U_A=0$ at $t=0$.

After the wave energy increases to $U_A\sim U_M$, where $U_M=B^2/(8\pi)$
is the energy of the background magnetic field, the waves are expected
to be damped by non-linear effects. Although the exact value of the
maximum of $U_A$ has not been known, we set it at $U_M$.  After the wave
energy increases to the maximum, the waves are expected to heat ICM
\citep[e.g.][]{ohira09,gargate10}. Thus, we give the heating term in
equation~(\ref{eq:eg}) by
\begin{equation}
 H_{\rm st} = \Gamma v_A\left|\frac{\partial P_c}{\partial r}\right|
\end{equation}
\citep{voe84,kan06}. We expect that $\Gamma\sim 1$ after the wave energy
increases to $U_A\sim U_M$. Thus, we simply give $\Gamma=U_A/U_M$ for
$U_A<U_M$ and $\Gamma=1$ after $U_A$ reaches $U_M$. Since $U_A$ rapidly
increases at the cluster core (see \S~\ref{sec:cr}), the results are not
much dependent on the definition of $\Gamma$ for $U_A<U_M$. Note that
\citet{guo08} assumed that $\Gamma$ is always one.

As the source of CRs, we primarily consider the AGN at the center of the
cluster. The CRs may be supplied from bubbles observed in the central
region of clusters \citep{guo08}. Moreover, they may also be supplied by
strong outbursts of the AGN. In this case, CRs are accelerated at the
forward shock of a cocoon and they are distributed in a broad region
\citep{fuj07c}. In fact, this kind of outbursts have been observed in
several clusters \citep{mcn05,nul05a,nul05b}. Furthermore, cluster
mergers generate turbulence around the core
\citep*{fuj04c,fuj05a,asc06}, and the turbulence could accelerate and
provide CRs in the central region of the clusters
\citep*[e.g.][]{ohn02,fuj03,bru04,cas05,ens11,bru11}. Although the
supply of the CRs may be intermittent, we study continuous CR
injection as the time-average. The source term we adopt is similar to
that of \citet{guo08} and it is simply given by
\begin{equation}
\label{eq:dotSc}
 \dot{S}_c = \frac{3-\nu}{4\pi}
\frac{L_{\rm AGN}}{r_1^3 (r_1/r_0)^{-\nu}-r_0^3}
\left(\frac{r}{r_0}\right)^{-\nu}(1-e^{-(r/r_0)^2})e^{-(r/r_1)^2}\:,
\end{equation}
where $L_{\rm AGN}$ is the energy injection rate from the AGN. This
means that CRs are mostly injected at $r_0\lesssim r\lesssim r_1$. We
adopt $r_0=20$~kpc following \citet{guo08} based on the observations of
bubbles. We also adopt $r_1=150$~kpc, which is the size of the radio
minihalo observed in the Perseus cluster \citep*{git02}. \citet{fuj07c}
indicated that the ICM might be heated within that radius. The injection
rate is $L_{\rm AGN}=\epsilon \dot{M} c^2$, where $\epsilon$ is the
parameter, $\dot{M}$ is the inflow rate of the gas toward the AGN, and
$c$ is the speed of light. Following \citet{guo08}, we assume $\nu\sim
3$.

The diffusion of a CR particle depends on its energy, because its
resonance with an Alfv\'en wave depends on the gyro radius. This means
that the diffusion coefficient $D$ depends on the energy spectrum of
CRs. However, the energy spectrum of CRs ejected from the AGN is
unknown. Therefore, we give it as a simple function of the ICM density:
\begin{equation}
 D(\rho) = D_0 (\rho/\rho_0)^{-d} \:,
\end{equation}
where $D_0$ and $\rho_0$ are respectively the values of $D$ and $\rho$
at $r=0$ and $t=0$, and $d$ is the parameter. We included the dependence
on $\rho$, because we expect that the diffusion coefficient is reduced as
the ICM is compressed and the magnetic fields are increased
\citep{mat09}. We assume that the magnetic fields are $B=B_0
(\rho/\rho_0)^{d}$, where $B_0$ is the parameter and we take $d=2/3$.

\subsection{Gravitational Matter and Gas Profile}

We consider two types of clusters. One is a relatively large cluster and
the other is a small cluster. For the profile of the acceleration of
gravity or the profile of gravitational matter, we adopt models
constructed on observations of the Perseus cluster (large) the Virgo
cluster (small).

For the Perseus cluster, \citet{mat06} gave an analytical profile of the
acceleration of gravity, $G M(r)/r^2$, which is constructed based on the
observations by \citet{chu04}. We adopt the profile including the
contribution of the central galaxy \citep[\S~2 in][]{mat06}. For the
Virgo cluster, we use the density and temperature profiles obtained by
\citet{ghi04}. From their results, we can construct the mass profile
assuming that the ICM is almost in pressure equilibrium.

We assume that the ICM is initially isothermal. This is because we do
not know the initial distribution of CRs. If we assume that the ICM is
isothermal, $u=0$, and $P_c=0$ at $t=0$, the ICM density is relatively
low at the cluster center, and $\dot{M}$ is small when $t$ is
small. Then, as the ICM cools at the cluster center, $\dot{M}$ and the
activity of the central AGN gradually increase, and the CRs injected by
the AGN are accumulated in the core. On the other hand, if we adopt the
observed current density and temperature profiles, the density at the
cluster center is high and the cooling time is small. If we start
calculations with $u=0$ and $P_c=0$ at $t=0$, $\dot{M}$ abruptly
increases and CRs are injected in a very short time, which causes
numerical instability. In reality, it is likely that AGN activities
precede cluster formation. Thus, it is natural to assume that some
amount of CRs had already been injected when the cool core was
established.

At $t=0$, we assume that the temperature of the large cluster is 7~keV
and that of the small cluster is 2.4~keV, which are the values in the
outer region of the Perseus and the Virgo cluster, respectively
\citep{chu04,ghi04}. The ICM profiles are built so that the ICM is in
pressure equilibrium in the given gravitational fields. The
normalization of the density is determined so that the density in the
outer region of the clusters is identical to the observed ones. The
initial velocity of the ICM is $u=0$. There are no CRs at $t=0$.

\section{Results}
\label{sec:res}

\subsection{Numerical Methods}

The hydrodynamic part of the equations is solved by a second-order
advection upstream splitting method (AUSM) based on \citet[][see also
\citealt*{wad01,fuj04a}]{lio93}. We use 300 unequally spaced meshes in
the radial coordinate to cover a region with a radius of 1~Mpc. The
inner boundary is set at $r_{\rm min}=5$~kpc. We adopt inflow/outflow
boundary conditions at the inner and outer radii. Model parameters are
presented in Table~\ref{tab:par}. For gravitational potential, 'P'
refers to the Perseus type cluster, and 'V' refers to the Virgo type
cluster.

\begin{figure}
\epsscale{.80}
\plotone{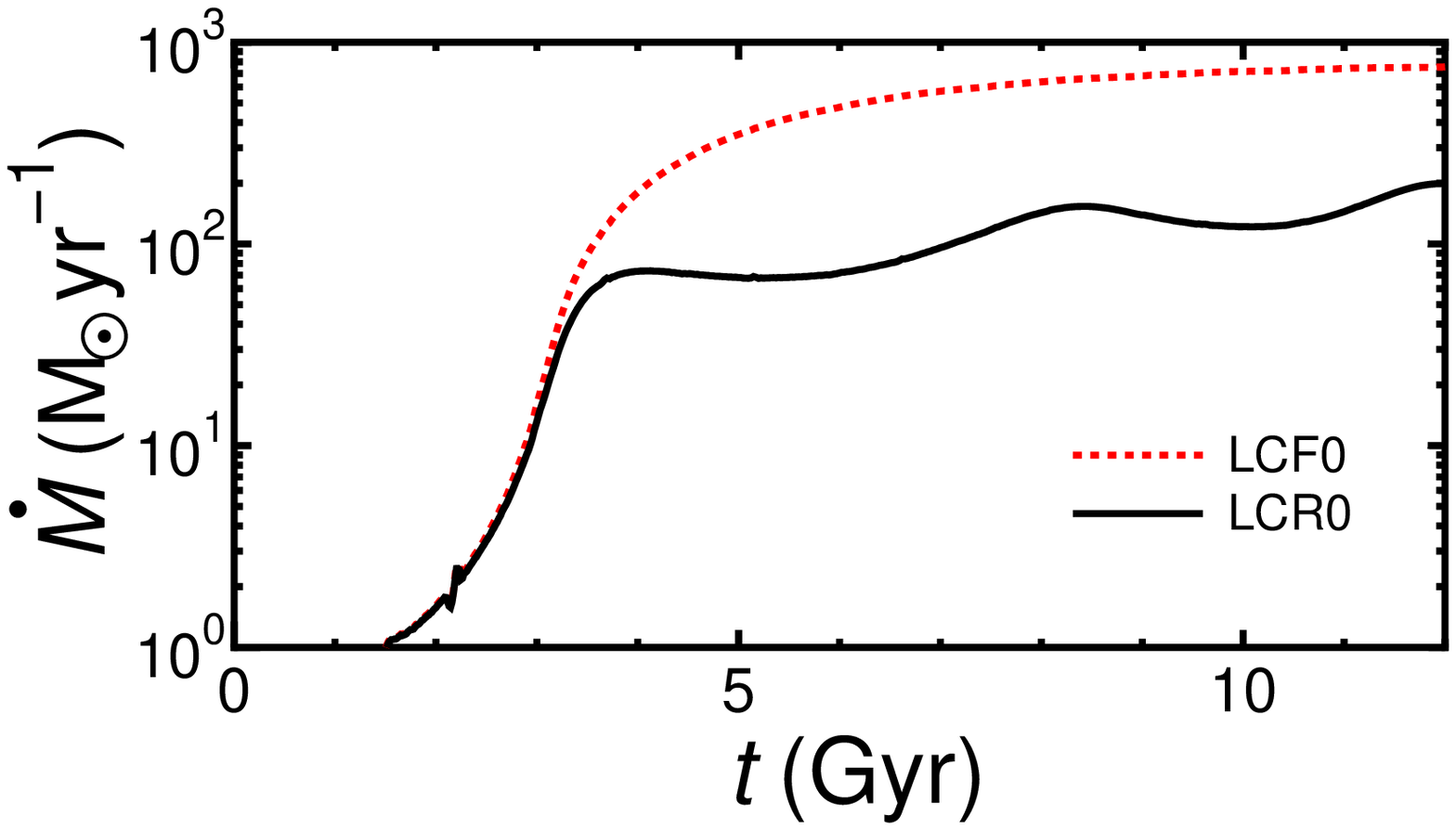}
\caption{Evolution of $\dot{M}$ for Models~LCF0, and LCR0}
\label{fig:dotM_lcf0}
\end{figure}

Before we investigate heating by CR streaming. we study a pure cooling
flow model for comparison. If there is no heating source ($\epsilon=0$
and $f_c=0$; Model LCF0), a cooling flow develops and reaches almost a
steady state at $t\gtrsim 4$~Gyr. Figure~\ref{fig:dotM_lcf0} shows the
evolution of $\dot{M}$. At $t=12$~Gyr, $\dot{M}$ increases to $760\:
M_\odot\:\rm yr^{-1}$.

\subsection{CR Heating}
\label{sec:cr}

In this subsection, we consider our fiducial model with CR heating
(Model~LCR0). We did not include thermal conduction or $f_c = 0$. The
initial magnetic field and the diffusion coefficient at the cluster
center are $B_0=10\:\mu$G and $D_0=1\times 10^{26}\rm\: cm^2\: s^{-1}$,
respectively. We use the same value of $D_0$ for other models including
CR heating. The diffusion coefficient is much smaller than the values
assumed by \citet{guo08} and \citet{mat09}. This is because the Alfv\'en
waves become non-linear ($\delta B\sim B$) as we show bellow. In this
case, the diffusion coefficient is close to the one for the Bohm
diffusion (equation~[14] in \citealt{bel78}) and can be very
small. Effectively, the diffusion coefficient we adopted is too small to
affect the results. In other words, the results are not much different
even if we assume $D_0=0$.

\begin{figure}
\epsscale{.80}
\plotone{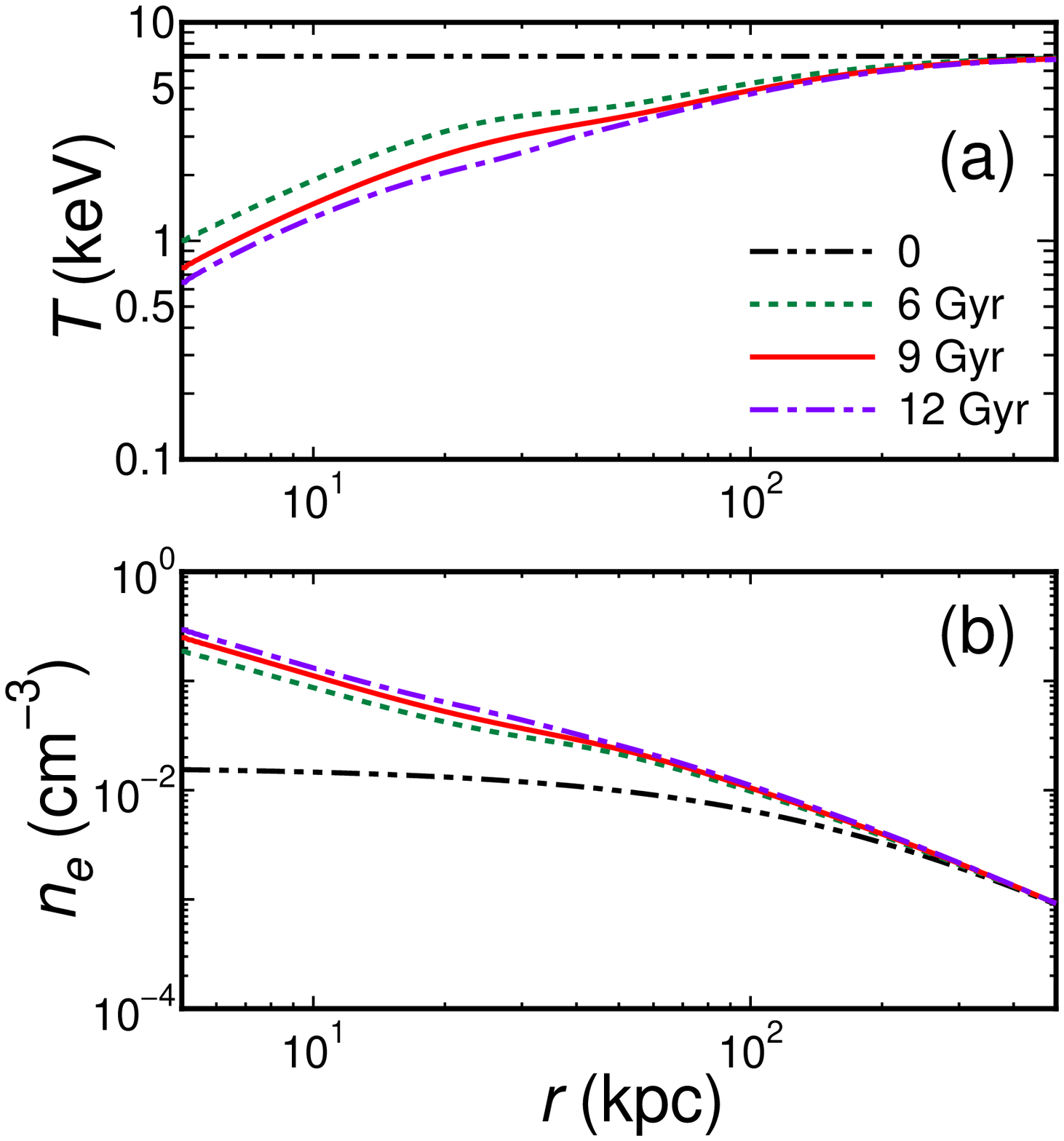}
\caption{(a) Temperature and (b) density profiles for Model~LCR0.}
\label{fig:Tn_lcr0}
\end{figure}

\begin{figure}
\epsscale{.80}
\plotone{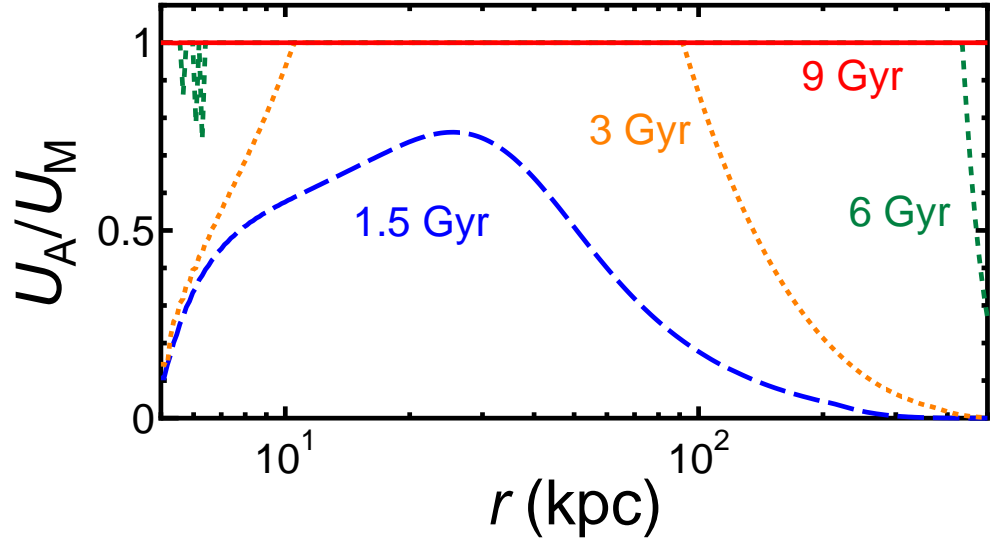}
\caption{Profiles of the ratio $U_A/U_M$ for Model~LCR0.}
\label{fig:UA_lcr0}
\end{figure}

Figure~\ref{fig:Tn_lcr0} shows the profiles of ICM temperature and
density for Model~LCR0. For $t\gtrsim 4$~Gyr, they do not much
change. However, the temperature of the inner boundary $r=r_{\rm min}$
slowly oscillates at $0.6\lesssim k_B T\lesssim 1$~keV. The oscillation
is reflected in $\dot{M}$ (Figure~\ref{fig:dotM_lcf0}). Compared with
the pure cooling flow model (Model~LCF0), $\dot{M}$ is significantly
reduced ($\sim 120\: M_\sun\:\rm yr^{-1}$ at $t\sim 12$~Gyr), which
means that CR streaming can be an effective, stable heating
source. Figure~\ref{fig:UA_lcr0} shows the evolution of the ratio
$U_A/U_M$. The region where $U_A/U_M$ reaches one expands inward and
outward in the cluster. This means that the Alfv\'en waves become
non-linear in a wide region of the cluster at the end of the
calculation. It is to be noted that even if we assume $U_A/U_M=1$
throughout the calculation, the results do not much change. The growth
time of $U_A$ is much larger than that for a supernova remnant. The main
reason is that the spatial scale of a cluster is much larger than that
of the precursor of the shock of a supernova remnant. The difference
affects the gradient of $P_c$ in equation (\ref{eq:UA}).

\begin{figure}
\epsscale{.80} 
\plotone{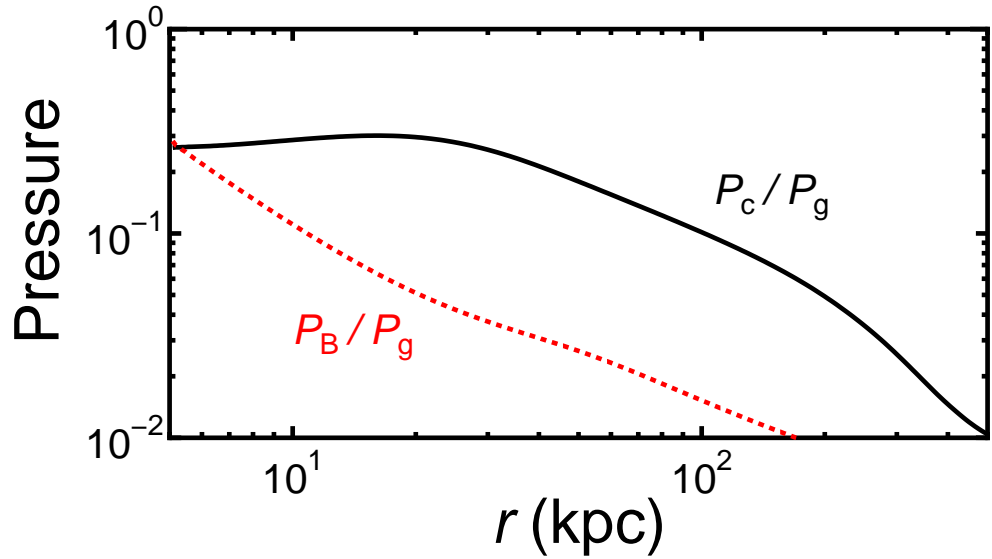} 
\caption{Profiles of the ratios
$P_c/P_g$ (solid) and $P_B/P_g$ (dotted) at $t=9$~Gyr for Model~LCR0.}
\label{fig:Pcb_lcr0}
\end{figure}

\begin{figure}
 \epsscale{.80}
 \plotone{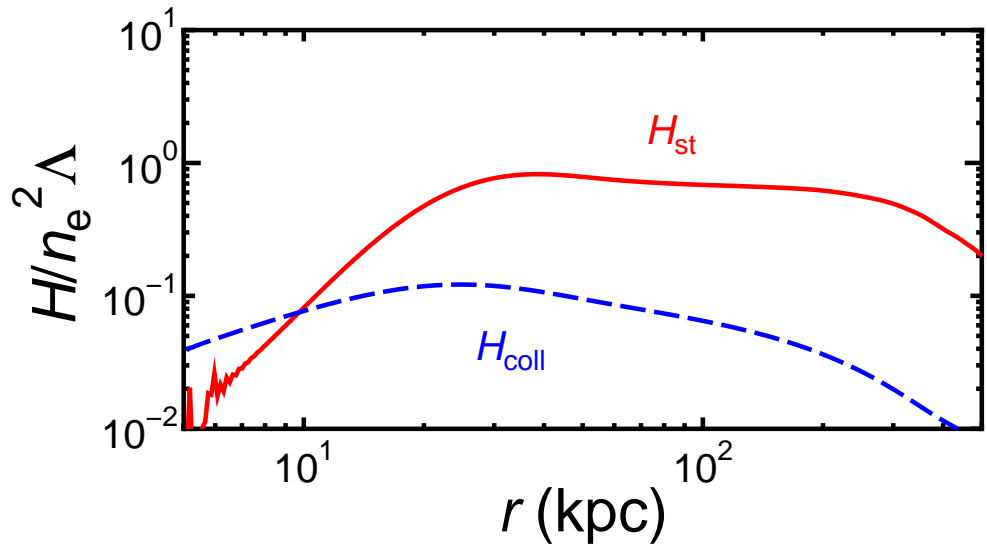}
 \caption{Relative importance of CR streaming ($H_{\rm st}$) and
 collisional heating ($H_{\rm coll}$)
at $t=9$~Gyr for Model~LCR0.}
\label{fig:heat_lcr0}
\end{figure}

\begin{figure}
\epsscale{.80}
\plotone{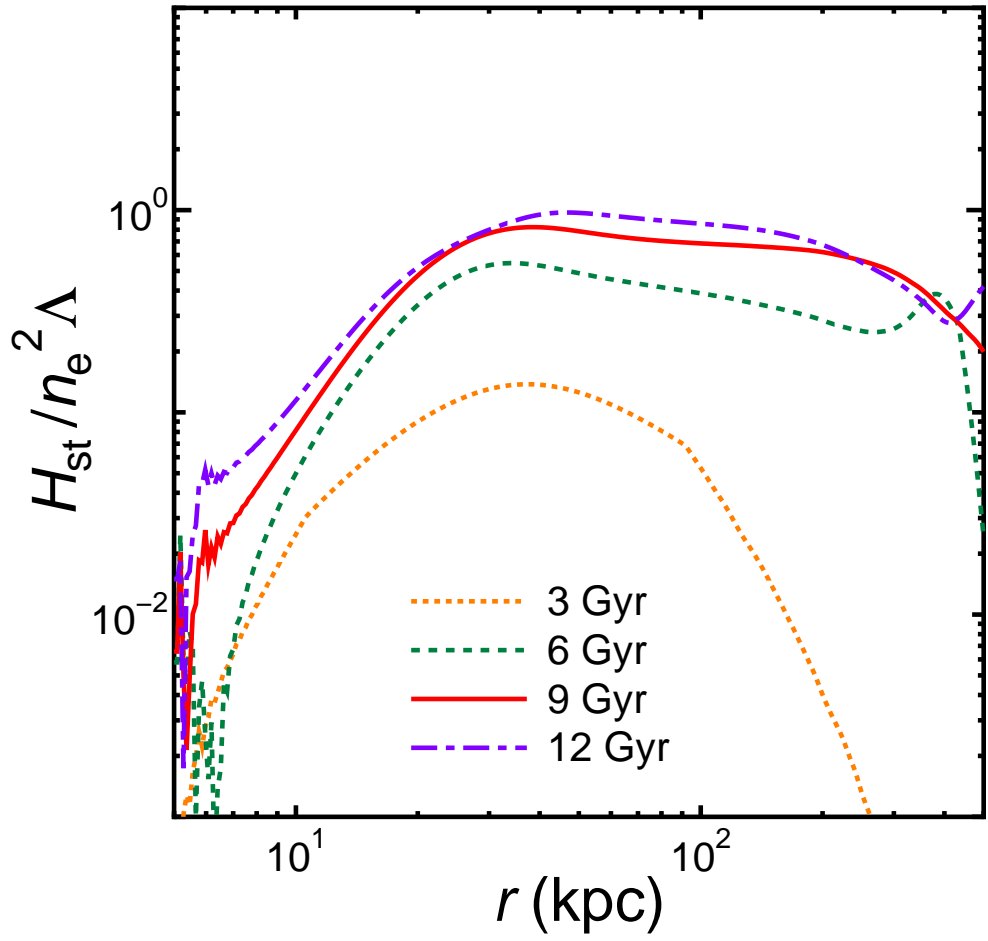}
\caption{Evolution of $H_{\rm st}/(n_e^2\Lambda)$ for Model~LCR0.}
\label{fig:heat2_lcr0}
\end{figure}

Figure~\ref{fig:Pcb_lcr0} shows the ratios $P_c/P_g$ and $P_B/P_g$ at
$t=9$~Gyr. Both the CR and magnetic pressures are smaller than the gas
pressure, although they are relatively large in the central region and
cannot be ignored ($P_c/P_g\sim 0.3$ and $P_B/P_g\sim 0.3$). In
Figure~\ref{fig:heat_lcr0}, we show relative importance of the two
heating mechanisms ($H_{\rm st}$ and $H_{\rm coll}$) at $t=9$~Gyr. CR
streaming alone can almost balance with radiative cooling except for the
very inner region of the cluster. This means that it can heat almost the
entire cool core of $r\lesssim 100$~kpc. The contribution of Coulomb and
hadronic collisions are minor.  Figure~\ref{fig:heat2_lcr0} shows the
evolution of the ratio of the heating by CR streaming to the radiative
cooling. The ratio gradually reaches one. The bend at $r\sim 400$~kpc at
$t=6$~Gyr corresponds to the point where $U_A$ reaches $U_M$
(Figure~\ref{fig:UA_lcr0}).

\subsection{Parameter Search}

\begin{figure}
\epsscale{.80} \plotone{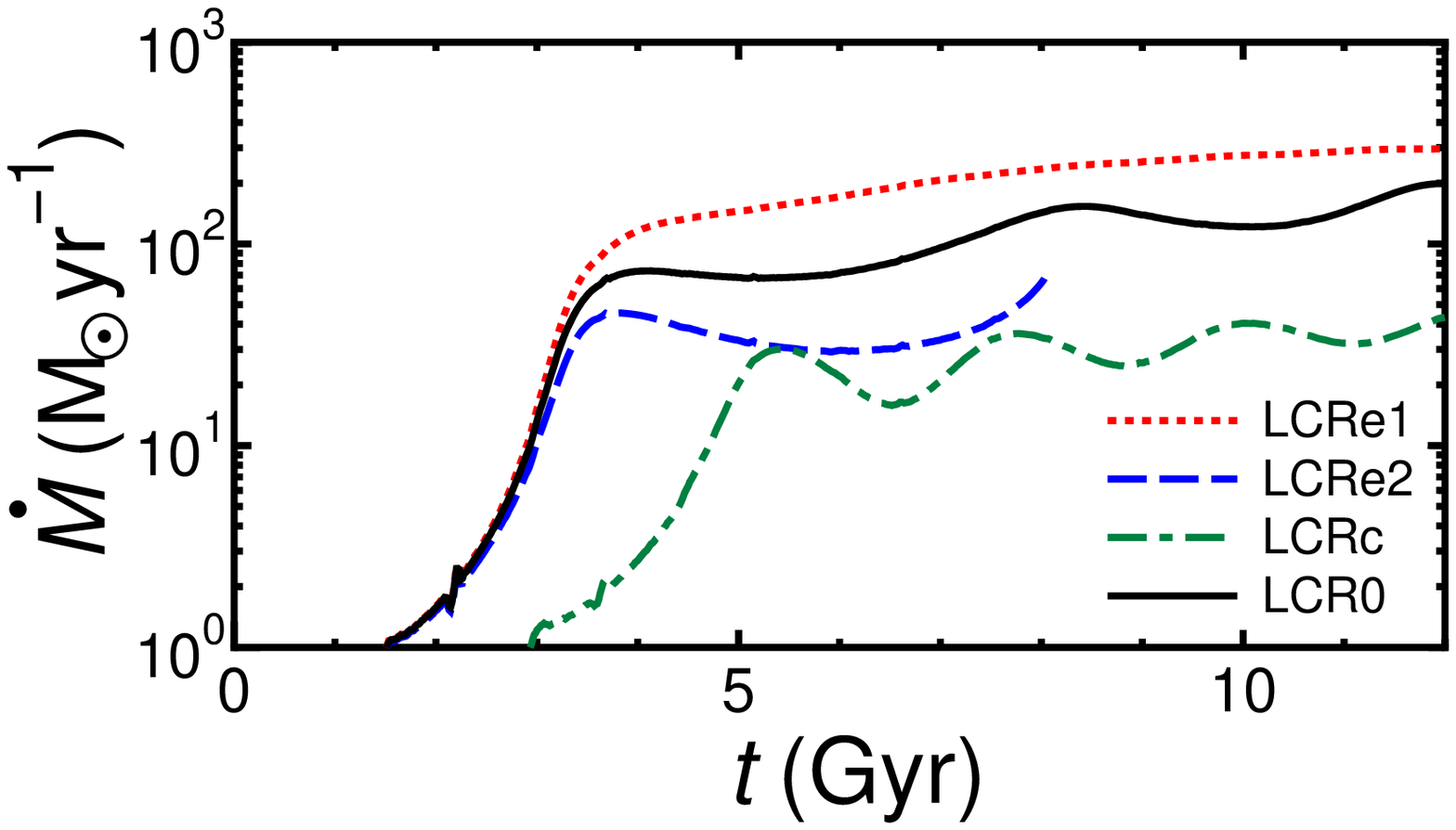} 
\caption{The evolution of $\dot{M}$ for Models~LCRe1, LCRe2, LCRc, 
and~LCR0.}  
\label{fig:dotM_lcre}
\end{figure}

\begin{figure}
\epsscale{.80}
\plotone{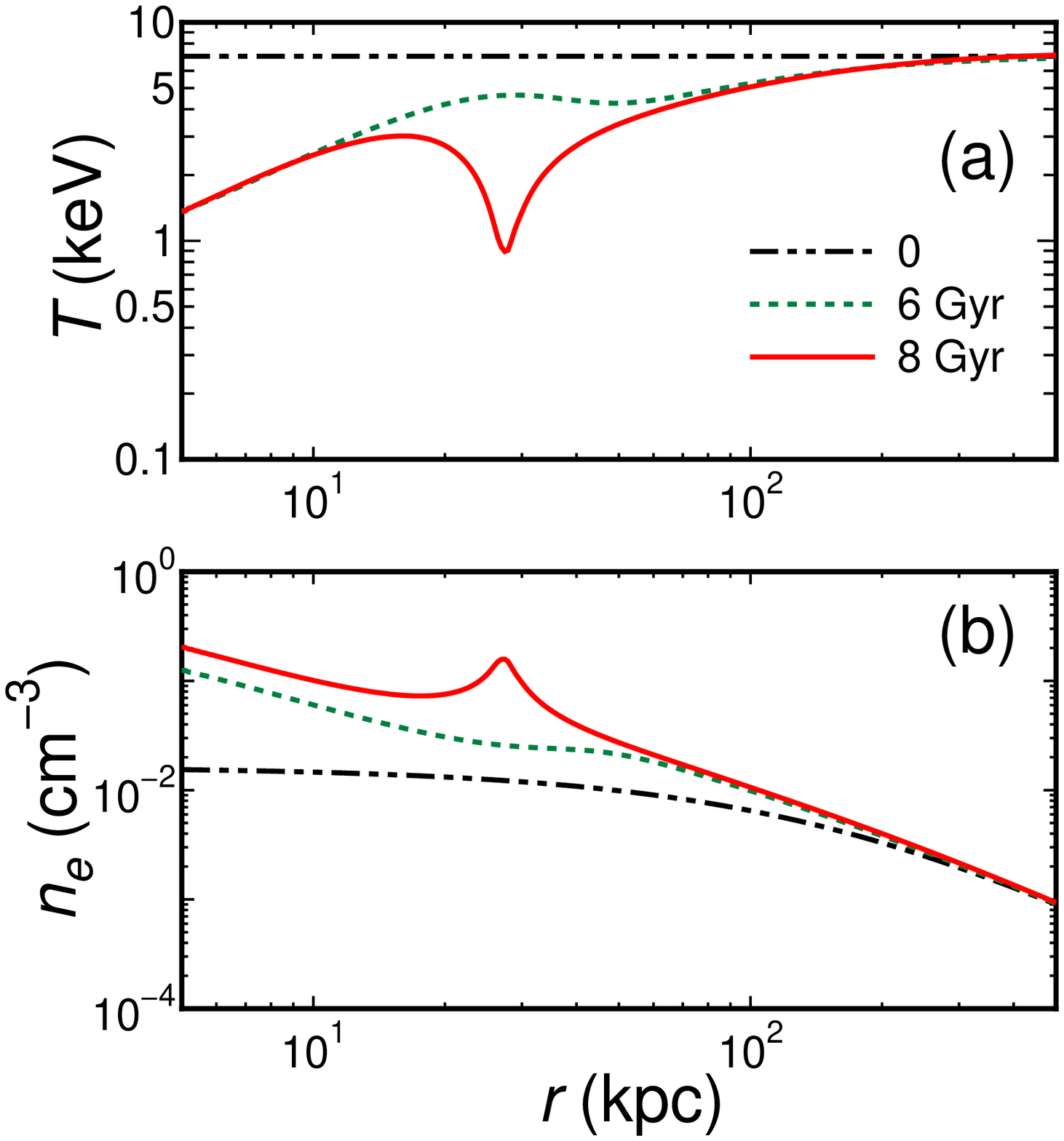}
\caption{(a) Temperature and (b) density profiles for Model~LCRe2.}
\label{fig:Tn_lcre2}
\end{figure}

In this subsection, we change model parameters to see how the results
are affected by them. Probably, most uncertain parameters in our models
are those for the energy input from the AGN. In Models~LCRe1 and LCRe2,
we change the value of $\epsilon$ (Table~\ref{tab:par}). The evolution
of $\dot{M}$ is presented in Figure~\ref{fig:dotM_lcre}. As is expected,
we tend to have a smaller $\dot{M}$ for a larger $\epsilon$. For
Model~LCRe2, the ICM becomes unstable for $t\gtrsim 8$~Gyr, although
$\dot{M}$ is smaller than that in Model~LCR0 for a long duration of
$\sim 5$~Gyr ($3\lesssim t\lesssim 8$~Gyr). The temperature and density
profiles are shown in Figure~\ref{fig:Tn_lcre2}. In our source model,
CRs are injected most intensively at $r\sim r_0$
(equation~[\ref{eq:dotSc}]). If $\epsilon$ is too large, radiative
cooling cannot cancel the CR heating at $r\sim r_0$, which makes the
temperature and density profiles irregular at $r\sim r_0$. In general,
models in which CRs are injected in a too narrow region and/or the ICM
is too strongly heated tend to be unstable. Since our simulations are
one-dimensional, we cannot investigate what happens after the ICM
distribution becomes irregular. Multi-dimensional simulations would be
interesting to study that. The instability for Model~LCRe2 can be
prevented by thermal conduction. Model~LCRc is the same as Model~LCRe2
but $f_c=0.1$. In this model, the ICM is stable even at $t\sim
12$~Gyr. Moreover, $\dot{M}$ is smaller than that in Model~LCR0
throughout the calculation and $\sim 40\: M_\sun\:\rm yr^{-1}$ at $t\sim
12$~Gyr (Figure~\ref{fig:dotM_lcre}). Note that if we do not include
heat sources except for the thermal conduction of $f_c=0.1$, the mass
inflow rate is $\dot{M}\sim 400\: M_\sun\:\rm yr^{-1}$ at $t\sim
12$~Gyr. This means that the thermal conduction of this level alone
cannot effectively halt a cooling flow.

\begin{figure}
\epsscale{.80}
\plotone{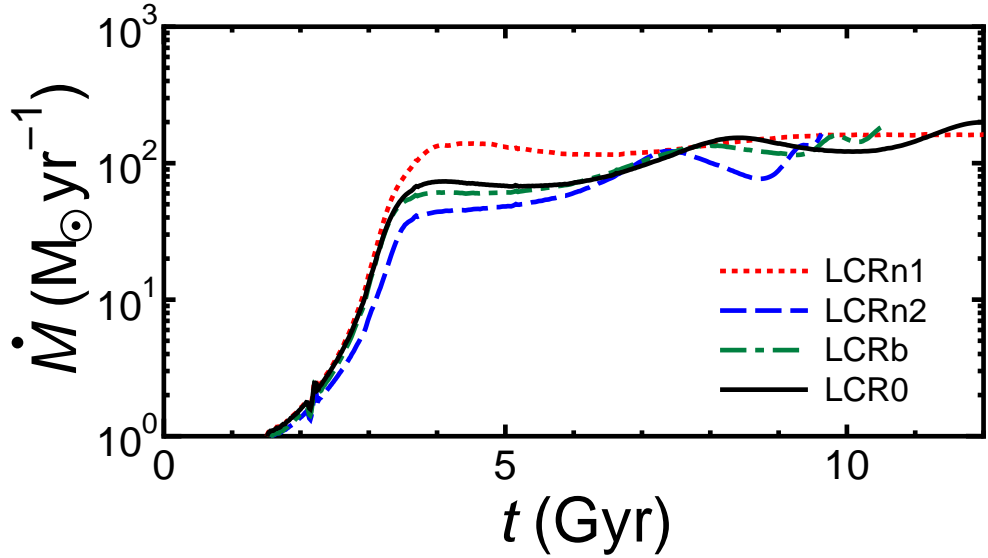}
\caption{Evolution of $\dot{M}$ for Models~LCRn1, LCRn2, LCRb, and~LCR0.}
\label{fig:dotM_lcrn}
\end{figure}

Models~LCRn1 and LCRn2 are the cases where the value of $\nu$ is changed
(Table~\ref{tab:par}). The evolution of $\dot{M}$ is presented in
Figure~\ref{fig:dotM_lcrn}. For a larger $\nu$, CRs are injected more
intensively at $r\sim r_0$ (equation~[\ref{eq:dotSc}]). Thus, the ICM
becomes unstable at $t\sim 9.6$~Gyr for Model~LCRn2. In Model~LCRb, we
change the strength of the background magnetic fields and adopt
$B_0=5\:\mu$G. The results are not much different from those for
Model~LCR0 ($B_0=10\:\mu$G; Figure~\ref{fig:dotM_lcrn}). However, the
ICM becomes unstable at $t\sim 10.5$~Gyr.

\begin{figure}
\epsscale{.80}
\plotone{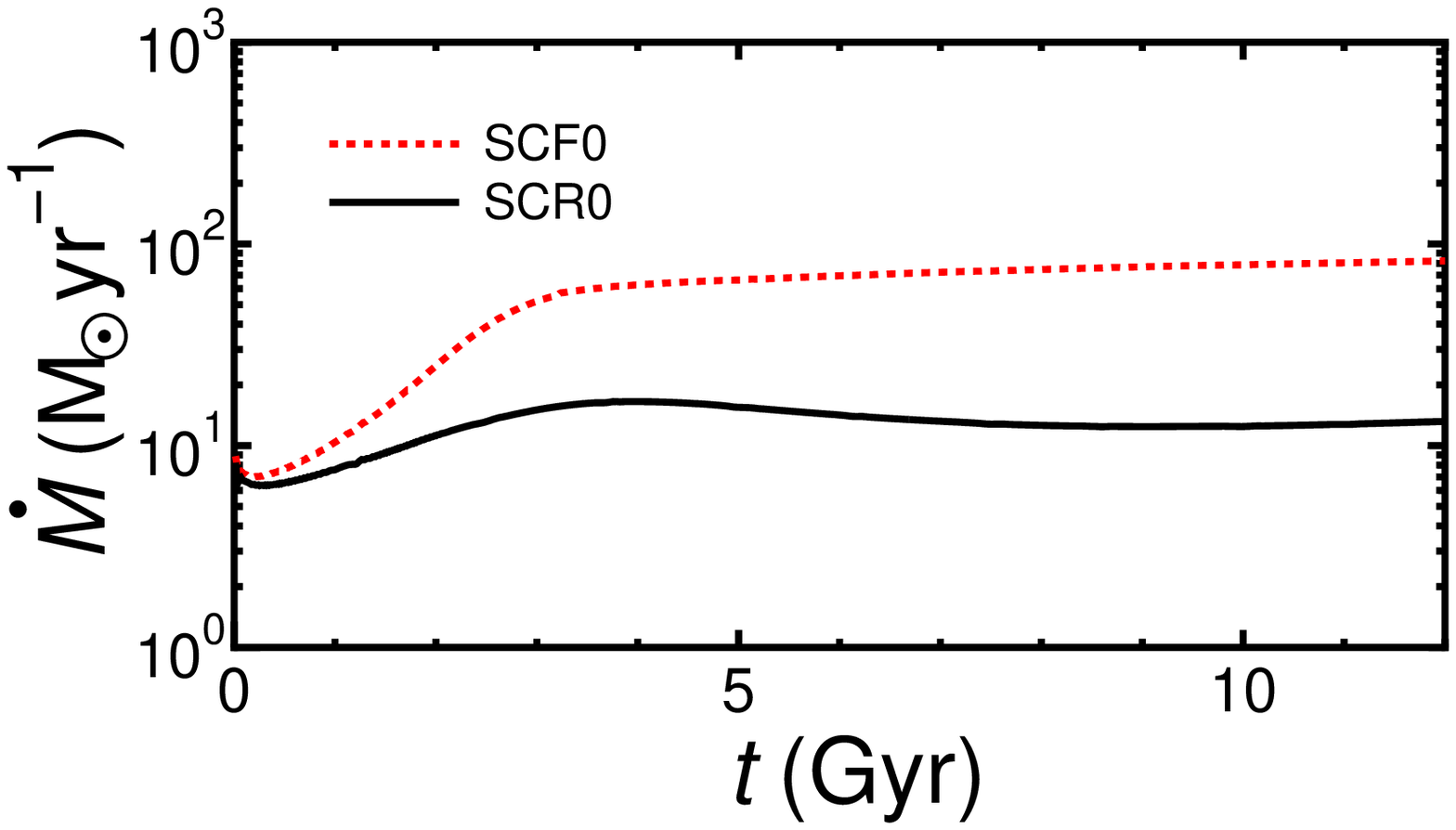}
\caption{Evolution of $\dot{M}$ for Models~SCF0, and SCR0}
\label{fig:dotM_s}
\end{figure}

\begin{figure}
\epsscale{.80}
\plotone{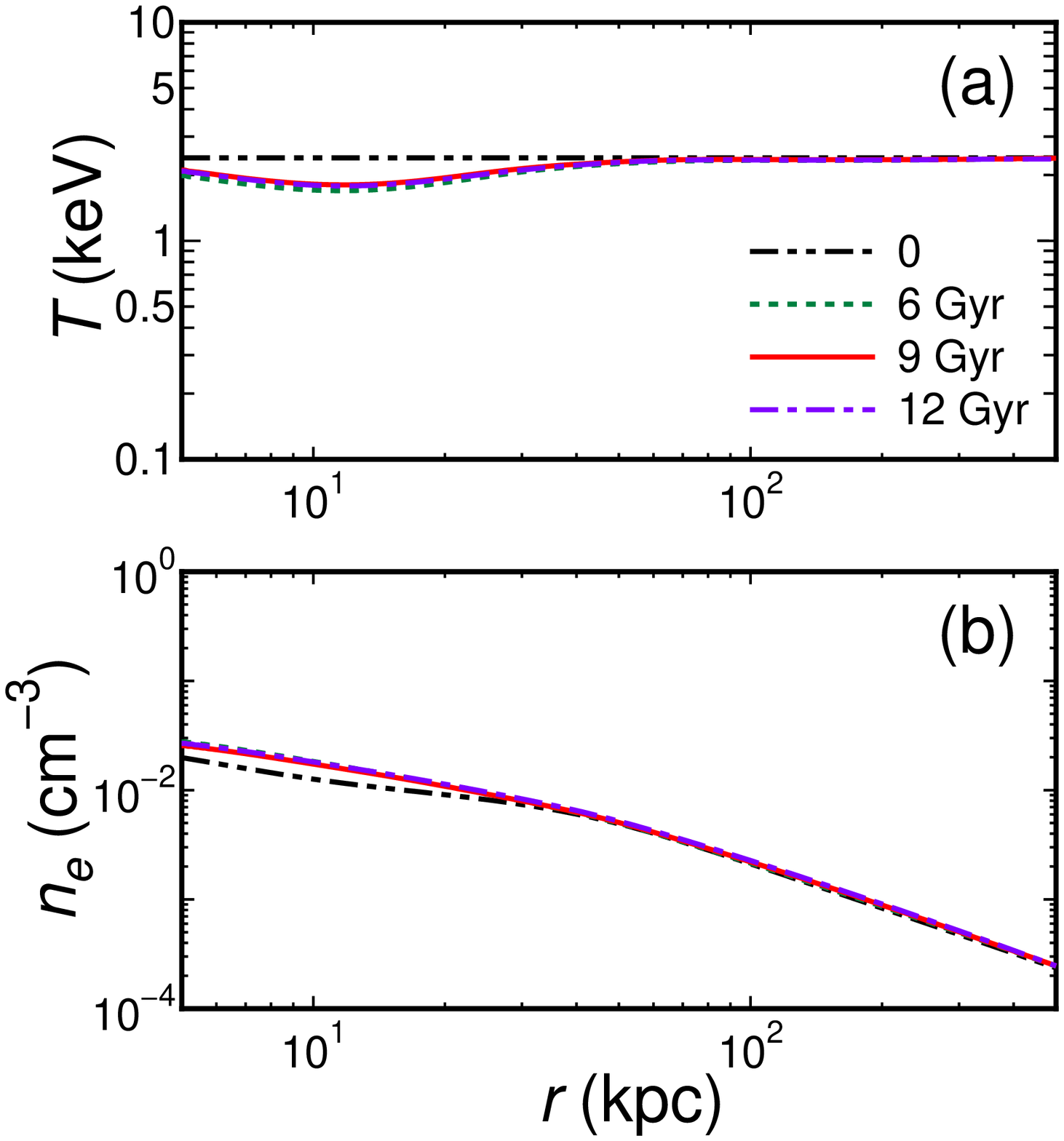}
\caption{(a) Temperature and (b) density profiles for Model~SCR0.}
\label{fig:Tn_scr0}
\end{figure}

We also considered a less massive cluster (the Virgo type). The initial
ICM temperature ($\sim 2.4$~keV) is much smaller than that of the
Perseus type cluster ($\sim 7$~keV). Model~SCF0 corresponds to a pure
cooling flow. In Figure~\ref{fig:dotM_s}, the mass flow rate amounts to
$\dot{M}\gtrsim 80\: M_\sun\:\rm yr^{-1}$ at the end of the
calculation. Model~SCR0 includes CR heating. The mass inflow rate is
significantly reduced by the heating and $\dot{M}\sim 13\: M_\sun\:\rm
yr^{-1}$ at $t\sim 12$~Gyr
(Figure~\ref{fig:dotM_s}). Figure~\ref{fig:Tn_scr0} shows the
temperature and density distributions for Model~SCR0. They are stable
until the end of the calculation even if there is no assistance of
thermal conduction.

\section{Discussion}

The above results show that heating by CR streaming can almost balance
with radiative cooling, and the heating process is relatively stable,
even if there is no thermal conduction. The stability can roughly be
explained as follows.

\begin{figure}
 \epsscale{.80} 
\plotone{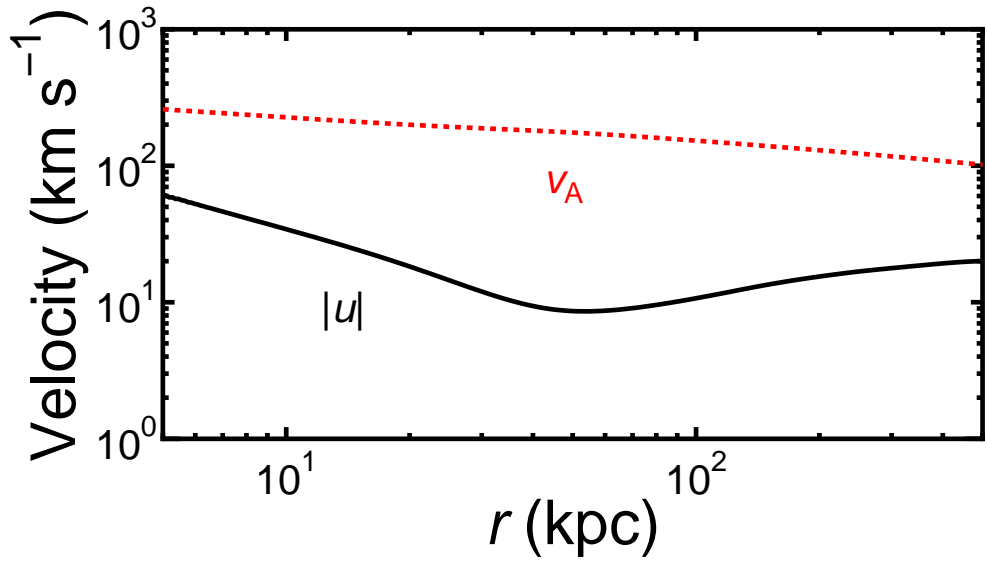} \caption{ICM velocity and Alfv\'en velocity 
profiles at $t=9$~Gyr for Model~LCR0.}
 \label{fig:v_lcr0}
\end{figure}

\begin{figure}
 \epsscale{.80} 
\plotone{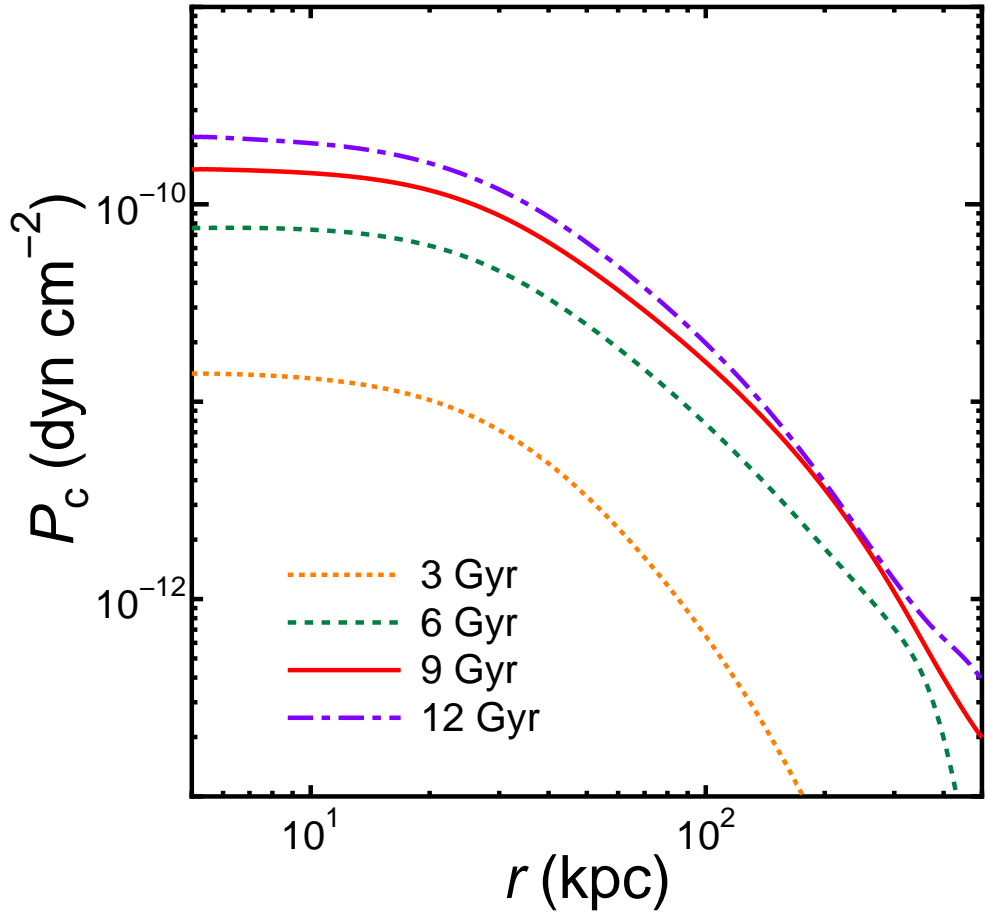} \caption{CR pressure profiles for Model~LCR0.}
 \label{fig:pcr2_lcr0}
\end{figure}

In our calculations, the ICM velocity is much smaller than the Alfv\'en
velocity. Figure~\ref{fig:v_lcr0} is an example (Model~LCR0 at
$t=9$~Gyr). This means that $v_A$ is the main contributer of
$\tilde{u}=u+v_A$ in equation~(\ref{eq:ec}). Owing to the large Alfv\'en
velocity, CRs can prevail in and heat the entire core, which is
different from other conventional heating mechanisms such as sound waves
or weak shocks \citep{fuj05,mat06}. Moreover, the Alfv\'en velocity is
given by $v_A=B/\sqrt{4\pi\rho}\propto \rho^{d-1/2}\propto \rho^{1/6}$
if we assume $d=2/3$. Thus, the Alfv\'en velocity is not much dependent
on the ICM. The diffusion term in equation~(\ref{eq:ec}) can effectively
be ignored because of the small value (\S~\ref{sec:cr}). These indicate
that the distribution of $P_c$ or $e_c$ is insensitive to the change of
local ICM, which makes the CR heating stable. Although the absolute
value of $P_c$ increases as $\dot{S}_c (\propto \dot{M})$ increases, the
overall shape of the profile of $P_c$ does not much change
(Figure~\ref{fig:pcr2_lcr0}). Moreover, since $P_c$ reflects
accumulation of CRs injected so far, it is insensitive to a temporal
change of $\dot{S}_c$.

The stability also resides in the heating function $H_{\rm
st}=v_A|\partial P_c/\partial r|$ and its global balance with the
radiative cooling function. Here, physical quantities are the typical
ones for $\lesssim r$. Since $v_A$ does not much evolve and the overall
shape of the profile of $P_c$ does not much change, the heating function
can be approximated by $H_{\rm st}\propto \dot{M}f(r)$, where $f(r)$ is
a function of $r$ and is almost independent of $t$. The mass inflow rate
is given by $\dot{M} = 4\pi r^2\rho |u|$, which does not depend on the
radius at a given time because of the mass conservation. The flow time
of the ICM, $t_{\rm flow}\propto r/|u|$, is nearly proportional to the
cooling time of the ICM, $t_{\rm cool}\propto P_g/(n_e^2 \Lambda)$,
because the flow compensates the cooled gas. Thus, we obtain $H_{\rm
st}\propto (\rho^3\Lambda/P_g) r^3 f(r)$. If the ICM is
adiabatic. $P_g\propto \rho^{\gamma_g}$, where $\gamma_g=5/3$. However,
radiative cooling is effective in the central region of a cluster, and
thus $P_g\propto \rho^{\gamma_g'}$, where
$1<\gamma_g'<5/3$. Observationally, $\gamma_g'=1.20\pm 0.06$ for
clusters with a cool core \citep{deg02}. Thus, the heating function
$H_{\rm st} (\propto \rho^{3-\gamma_g'}\Lambda)$ is similar to the
cooling term ($\propto \rho^2 \Lambda$) in equation
(\ref{eq:eg}). Because of this, the balance between the heating and the
cooling can be maintained.

However, the CR streaming is not perfectly locally stable as was noted
by \citet{loe91} (see their \S~2). Because of this, the ICM becomes
unstable at the end of calculations in some models without thermal
conduction (e.g. Models~LCRe2 and LCRn2). The local instability may be
related to emission-line filaments, which may be heated by CRs
\citep{loe91}. Even so, the ability to keep the ICM stable for a long
time makes CR streaming attractive as a heat source of cluster
cores. Moreover, the ability makes it easier for the model to attain
more stability when it is combined with minor thermal conduction
(Models~LCRc). Weak turbulence may also stabilize the ICM, because it
conveys energy as thermal conduction does \citep{kim03,fuj04c}.

\section{Conclusions}

We have studied heating of cool cores of galaxy clusters by CR
streaming. As the source of CRs, we considered the central AGN in a
cluster. The CRs amplify Alfv\'en waves, with which CRs move outward in
the cluster. The ICM is heated through dissipation of the waves.

Using numerical simulations, we found that CR streaming can heat the
core for a long time after radiative cooling becomes effective without
assistance of thermal conduction. Development of a strong cooling flow
is well prevented. CR streaming can effectively heat both high and low
temperature clusters. This is because CRs can prevail throughout the
core and their distribution is insensitive to the change of the
ICM. Thus, the entire core is heated by the CRs. Minor contribution of
thermal conduction makes the ICM even more stable.

There are limitations in our simple models. For example, we assumed
spherical symmetry of a cluster. In reality, it is likely that CRs are
injected anisotropically. Whether this anisotropy is erased or not
during the propagation of the CRs in the ICM may depend on the geometry
of the magnetic fields on which waves and CRs propagate. Moreover, the
CR injection or the acceleration itself is highly uncertain ($\dot{S}_c$
in equation~[\ref{eq:dotSc}]). In the future, $\gamma$-ray Observations
may reveal CR spectra, which may give us information on CR acceleration
in clusters.

\acknowledgments

We thank the anonymous referee for useful comments. This work was
supported by KAKENHI (Y.~F.: 20540269, 23540308).

\clearpage

\clearpage

\begin{deluxetable}{cccccc}
\tablecaption{Model Parameters \label{tab:par}}
\tablewidth{0pt}
\tablehead{
\colhead{Model} & Potential & $B_0$ ($\mu$G) 
& $f_c$ & $\epsilon$ & $\nu$ }
\startdata
LCF0   & P & 0  & 0   & 0                   & \nodata \\
LCR0   & P & 10 & 0   & $2.5\times 10^{-4}$ & 3.1     \\
LCRe1  & P & 10 & 0   & $1\times 10^{-4}$   & 3.1     \\
LCRe2  & P & 10 & 0   & $5\times 10^{-4}$   & 3.1     \\
LCRc   & P & 10 & 0.1 & $5\times 10^{-4}$   & 3.1     \\
LCRn1  & P & 10 & 0   & $2.5\times 10^{-4}$ & 2.5     \\
LCRn2  & P & 10 & 0   & $2.5\times 10^{-4}$ & 3.5     \\
LCRb   & P & 5  & 0   & $2.5\times 10^{-4}$ & 3.1     \\
SCF0   & V & 0  & 0   & 0                   & \nodata \\
SCR0   & V & 10 & 0   & $1\times 10^{-4}$   & 3.1     \\
\enddata
\end{deluxetable}


\begin{thebibliography}{}

\bibitem[Ascasibar 
\& Markevitch(2006)]{asc06} Ascasibar, Y., \& Markevitch,
		M.\ 2006, \apj, 650, 102 

\bibitem[Basson 
\& Alexander(2003)]{bas03} Basson, J.~F., \& Alexander, P.\ 2003,
		\mnras, 339, 353

\bibitem[Bell(1978)]{bel78} Bell, A.~R.\ 1978, \mnras, 182, 
147 

\bibitem[Berezhko 
\& Ellison(1999)]{ber99} Berezhko, E.~G., \& Ellison,
		D.~C.\ 1999, \apj, 526, 385 

\bibitem[Blanton et al.(2003)Blanton, Sarazin, \& McNamara]{bla03}
		Blanton, E.~L.,
Sarazin, C.~L., \& McNamara, B.~R.\ 2003, \apj, 585, 227 

\bibitem[Blanton et al.(2001)]{bla01} 
Blanton, E.~L., Sarazin, C.~L., McNamara, B.~R., \& Wise, M.~W.\ 2001, 
\apjl, 558, L15 

\bibitem[B\"ohringer 
\& Morfill(1988)]{boh88} B\"ohringer, H., \& Morfill, G.~E.\ 1988, \apj,
		330, 609

\bibitem[Bregman \& David(1988)]{bre88} Bregman, J.~N.,~\& 
David, L.~P.\ 1988, \apj, 326, 639 

\bibitem[Brighenti \& Mathews(2003)]{bri03} Brighenti, F.,~\& 
Mathews, W.~G.\ 2003, \apj, 587, 580 

\bibitem[Br{\"u}ggen 
\& Kaiser(2002)]{bru02} Br{\"u}ggen, M., \& Kaiser, C.~R.\ 2002, \nat,
		418, 301

\bibitem[Brunetti et al.(2004)]{bru04} Brunetti, G., Blasi, 
P., Cassano, R., \& Gabici, S.\ 2004, \mnras, 350, 1174 

\bibitem[Brunetti 
\& Lazarian(2011)]{bru11} Brunetti, G., \& Lazarian, A.\
		2011, \mnras, 412, 817 

\bibitem[Caprioli et al.(2009)Caprioli, Blasi, \& Amato]{cap09b}
		Caprioli, D., Blasi,
P., \& Amato, E.\ 2009, \mnras, 396, 2065 

\bibitem[Cassano 
\& Brunetti(2005)]{cas05} Cassano, R., \& Brunetti, G.\
		2005, \mnras, 357, 1313 

\bibitem[Churazov et al.(2001)]{chu01} Churazov, E., 
Br{\"u}ggen, M., Kaiser, C.~R., B{\"o}hringer, H., 
\& Forman, W.\ 2001, \apj, 554, 261 

\bibitem[Churazov et al.(2004)]{chu04} Churazov, E.,
		Forman,
W., Jones, C., Sunyaev, R., B\"ohringer, H.\ 2004, \mnras, 347, 29 

\bibitem[Colafrancesco et 
al.(2004)Colafrancesco, Dar, \& De R{\'u}jula]{col04} Colafrancesco, S.,
		Dar, A., \& De R{\'u}jula, A.\ 2004, \aap, 413, 441

\bibitem[De Grandi 
\& Molendi(2002)]{deg02} De Grandi, S., \& Molendi, S.\
		2002, \apj, 567, 163 

\bibitem[En{\ss}lin et 
al.(2011)]{ens11} En{\ss}lin, T., Pfrommer, C., Miniati, F., \&
		Subramanian, K.\ 2011, \aap, 527, A99

\bibitem[Ettori et al.(2002)]{ett02} Ettori, S., Fabian, 
A.~C., Allen, S.~W., \& Johnstone, R.~M.\ 2002, \mnras, 331, 635 

\bibitem[Ettori 
\& Fabian(2000)]{ett00} Ettori, S., \& Fabian, A.~C.\
		2000, \mnras, 317, L57 

\bibitem[Fabian(1994)]{fab94} Fabian, A.~C.\ 1994, \araa,
		32, 277 

\bibitem[Fabian et al.(2000)]{fab00} Fabian, A.~C.~et al.\ 
2000, \mnras, 318, L65 

\bibitem[Fabian et al.(2003)]{fab03a} Fabian, A.~C., Sanders, 
J.~S., Allen, S.~W., Crawford, C.~S., Iwasawa, K., Johnstone, R.~M., 
Schmidt, R.~W., \& Taylor, G.~B.\ 2003, \mnras, 344, L43 

\bibitem[Forman et al.(2005)]{for05} Forman, W., et al.\ 
2005, \apj, 635, 894 

\bibitem[Fujita et al.(2007a)]{fuj07c} Fujita, Y., Kohri, K., 
Yamazaki, R., \& Kino, M.\ 2007a, \apjl, 663, L61 


\bibitem[Fujita et al.(2004a)Fujita, Matsumoto, \& Wada]{fuj04c} Fujita,
		Y., Matsumoto,
T., \& Wada, K.\ 2004a, \apjl, 612, L9 

\bibitem[Fujita et al.(2005)]{fuj05a} Fujita, Y., Matsumoto, 
T., Wada, K., \& Furusho, T.\ 2005, \apjl, 619, L139 

\bibitem[Fujita et al.(2003)Fujita, Takizawa, \& Sarazin]{fuj03} Fujita,
		Y., Takizawa,
M., \& Sarazin, C.~L.\ 2003, \apj, 584, 190 

\bibitem[Fujita et al.(2002)]{fuj02} Fujita, Y., Sarazin, 
C.~L., Kempner, J.~C., Rudnick, L., Slee, O.~B., Roy, A.~L., 
Andernach, H., 
\& Ehle, M.\ 2002, \apj, 575, 764 

\bibitem[Fujita et al.(2004b)]{fuj04} Fujita, Y., Sarazin,
C.~L., Reiprich, T.~H., Andernach, H., Ehle, M., Murgia, M., 
Rudnick, L., 
\& Slee, O.~B.\ 2004b, \apj, 616, 157 

\bibitem[Fujita 
\& Suzuki(2005)]{fuj05} Fujita, Y., \& Suzuki, T.~K.\
		2005, \apjl, 630, L1 

\bibitem[Fujita et al.(2007b)]{fuj07b} Fujita, Y., Suzuki, 
T.~K., Kudoh, T., \& Yokoyama, T.\ 2007b, \apjl, 659, L1 

\bibitem[Fujita et al.(2004c)Fujita, Suzuki, \& Wada]{fuj04a} Fujita, Y.,
		Suzuki,
T.~K., \& Wada, K.\ 2004c, \apj, 600, 650 

\bibitem[Gargat\'{e} et al.(2010)]{gargate10} Gargat\'{e}, L., Fonseca,
		R.
 A., Niemiec, J., Pohl, M., Bingham, R., \& Silva, L. O., 2010, \apj, 
711, L127

\bibitem[Ghizzardi et al.(2004)]{ghi04} Ghizzardi, S., 
Molendi, S., Pizzolato, F., \& De Grandi, S.\ 2004, \apj, 609, 638 

\bibitem[Gitti et 
al.(2002)Gitti, Brunetti, \& Setti]{git02} Gitti, M., Brunetti, G., \&
		Setti, G.\
	  2002, \aap, 386, 456 

\bibitem[Guo 
\& Oh(2008)]{guo08} Guo, F., \& Oh, S.~P.\ 2008, \mnras,
	    384, 251 

\bibitem[Ikebe et al.(1997)]{ike97} Ikebe, Y., et al.\ 1997, 
\apj, 481, 660 

\bibitem[Johnstone et al.(2002)]{joh02} Johnstone, R.~M., 
Allen, S.~W., Fabian, A.~C., \& Sanders, J.~S.\ 2002, \mnras, 336, 299 

\bibitem[Jubelgas et 
al.(2008)]{jub08} Jubelgas, M., Springel, V., En{\ss}lin, T., \&
		Pfrommer, C.\ 2008, \aap, 481, 33

\bibitem[Kaastra et 
al.(2001)]{kaa01} Kaastra, J.~S., Ferrigno, C., Tamura, T., Paerels,
		F.~B.~S., Peterson, J.~R., \& Mittaz, J.~P.~D.\ 2001,
		\aap, 365, L99

\bibitem[Kang 
\& Jones(2006)]{kan06} Kang, H., \& Jones, T.~W.\ 2006, Astroparticle
		Physics, 25, 246

\bibitem[Kempner et al.(2002)Kempner, Sarazin, \&
			       Ricker]{kem02} Kempner,
J.~C., Sarazin, C.~L., \& Ricker, P.~M.\ 2002, \apj, 579, 236 

\bibitem[Kim 
\& Narayan(2003)]{kim03} Kim, W.-T., \& Narayan, R.\ 2003, \apjl, 596,
		L139

\bibitem[Kitayama 
\& Suto(1996)]{kit96b} Kitayama, T., \& Suto, Y.\ 1996,
	      \apj, 469, 480 

\bibitem[Liou \& Steffen(1993)]{lio93} Liou,~M., \& Steffen,~C.\ 1993,
			       J. Comp. Phys., 107, 23

\bibitem[Loewenstein et al.(1991)Loewenstein, Zweibel, \&
		Begelman]{loe91} Loewenstein, M.,
Zweibel, E.~G., \& Begelman, M.~C.\ 1991, \apj, 377, 392 

\bibitem[Lucek 
\& Bell(2000)]{luc00} Lucek, S.~G., \& Bell, A.~R.\ 2000, \mnras, 314,
	      65

\bibitem[Makishima et al.(2001)]{mak01} Makishima, K., et 
al.\ 2001, \pasj, 53, 401 

\bibitem[Mathews(2009)]{mat09} Mathews, W.~G.\ 2009, \apjl, 
695, L49 

\bibitem[Mathews et al.(2006)Mathews, Faltenbacher, \& Brighenti]{mat06}
		Mathews, W.~G.,
Faltenbacher, A., \& Brighenti, F.\ 2006, \apj, 638, 659 

\bibitem[Matsushita et 
al.(2002)]{mat02} Matsushita, K., Belsole, E., Finoguenov, A.,
		B\"ohringer, H.\ 2002, \aap, 386, 77

\bibitem[Mazzotta et al.(2002)]{maz02} Mazzotta, P., Kaastra, 
J.~S., Paerels, F.~B., Ferrigno, C., Colafrancesco, S., Mewe, R., \& 
Forman, W.~R.\ 2002, \apjl, 567, L37 

\bibitem[McNamara et al.(2000)]{mcn00} McNamara, B.~R., et 
al.\ 2000, \apjl, 534, L135 

\bibitem[McNamara et al.(2001)]{mcn01} McNamara, B.~R.~et 
al.\ 2001, \apjl, 562, L149 

\bibitem[McNamara et al.(2005)]{mcn05} McNamara, B.~R., 
Nulsen, P.~E.~J., Wise, M.~W., Rafferty, D.~A., Carilli, C., Sarazin, 
C.~L., \& Blanton, E.~L.\ 2005, \nat, 433, 45 

\bibitem[Nulsen et al.(2005a)]{nul05a} Nulsen, P.~E.~J., 
McNamara, B.~R., Wise, M.~W., \& David, L.~P.\ 2005, \apj, 628, 629 


\bibitem[Nulsen et al.(2005b)]{nul05b} Nulsen, P.~E.~J., 
Hambrick, D.~C., McNamara, B.~R., Rafferty, D., Birzan, L., Wise, M.~W., 
\& David, L.~P.\ 2005, \apjl, 625, L9 

\bibitem[Ohira et al.(2009)]{ohira09} Ohira, Y., Reville, B., Kirk,
		J. G.,
 \& Takahara, F. 2009, \apj, 698, 445

\bibitem[Ohno et al.(2002)Ohno, Takizawa, \& Shibata]{ohn02} Ohno, H.,
		Takizawa, M.,
\& Shibata, S.\ 2002, \apj, 577, 658 

\bibitem[Peterson et 
al.(2001)]{pet01} Peterson, J.~R., et al.\ 2001, \aap,
	  365, L104 

\bibitem[Pfrommer et al.(2007)]{pfr07} Pfrommer, C., 
En{\ss}lin, T.~A., Springel, V., Jubelgas, M., 
\& Dolag, K.\ 2007, \mnras, 378, 385 

\bibitem[Quilis et al.(2001)Quilis, Bower, \& Balogh]{qui01} Quilis, V.,
		Bower,
R.~G., \& Balogh, M.~L.\ 2001, \mnras, 328, 1091 

\bibitem[Ruszkowski \& Begelman(2002)]{rus02} Ruszkowski, 
M.~\& Begelman, M.~C.\ 2002, \apj, 581, 223 

\bibitem[Rephaeli(1987)]{rep87} Rephaeli, Y.\ 1987, \mnras, 
225, 851 

\bibitem[Rephaeli 
\& Silk(1995)]{rep95} Rephaeli, Y., \& Silk, J.\ 1995, \apj, 442, 91 

\bibitem[Sarazin(1986)]{sar86} Sarazin, C.~L.\ 1986, Reviews 
of Modern Physics, 58, 1 

\bibitem[Skilling(1975)]{ski75} Skilling, J.\ 1975, \mnras, 
173, 255 

\bibitem[Soker(2003)]{sok03} Soker, N.\ 2003, \mnras, 342, 
463 

\bibitem[Sutherland \& Dopita(1993)]{sut93} Sutherland, 
R.~S.~\& Dopita, M.~A.\ 1993, \apjs, 88, 253 

\bibitem[Takahara 
\& Takahara(1979)]{tak79} Takahara, M., \& Takahara, F.\
		1979, Progress of Theoretical Physics, 62, 1253 

\bibitem[Takahara 
\& Takahara(1981)]{tak81} Takahara, M., \& Takahara, F.\
		1981, Progress of Theoretical Physics, 65, 369 

\bibitem[Takizawa et al.(2003)]{tak03} Takizawa, M., Sarazin, 
C.~L., Blanton, E.~L., \& Taylor, G.~B.\ 2003, \apj, 595, 142 

\bibitem[Tamura et 
al.(2001)]{tam01} Tamura, T., et al.\ 2001, \aap, 365, L87 

\bibitem[Tucker 
\& Rosner(1983)]{tuc83} Tucker, W.~H., \& Rosner, R.\
		1983, \apj, 267, 547 

\bibitem[V\"olk et 
al.(1984)V\"olk, Drury, \& McKenzie]{voe84} V\"olk, H.~J., Drury, L.~O.,
		\& McKenzie,
	  J.~F.\ 1984, \aap, 130, 19 

\bibitem[Wada \& Norman(2001)]{wad01} Wada, K.~\& Norman, 
C.~A.\ 2001, \apj, 547, 172 

\bibitem[Zakamska 
\& Narayan(2003)]{zak03} Zakamska, N.~L., \& Narayan, R.\ 2003, \apj,
		582, 162


\end{thebibliography}
\end{document}